%% file: main.tex
\documentclass[10pt, journal]{IEEEtran}

\usepackage{todonotes}
\usepackage{threeparttable}
\usepackage{tabularray}
\usepackage{textcomp}

\usepackage{multirow}
\usepackage{balance}
\usepackage{fancyhdr}

\usepackage[table,xcdraw]{xcolor}
\usepackage{hhline}
\usepackage{makecell} 
\usepackage{pifont}
\usepackage[hidelinks]{hyperref}

\usepackage{algorithm}
\usepackage{algpseudocode}
\usepackage{multirow}
\usepackage{amsmath}
\usepackage{float}
\usepackage{booktabs}

\fancypagestyle{firstpage}{%
\lhead{}
\rhead{}
\fancyhead[CO]{Accepted for publication at IEEE Journal On Emerging and Selected Topics In Circuits and Systems. DOI 10.1109/JETCAS.2026.3707339}
}

\newcommand{\green}[1]{{\color{black}#1}}
\newcommand{\blue}[1]{{\color{black}#1}}
\newcommand{\orange}[1]{{\color{black}#1}}
\newcommand{\purple}[1]{{\color{black}#1}}
\newcommand{\black}[1]{{\color{black}#1}}

\def\journalname{IEEE Journal On Emerging and Selected Topics In Circuits and Systems}

\def\BibTeX{{\rm B\kern-.05em{\sc i\kern-.025em b}\kern-.08em
    T\kern-.1667em\lower.7ex\hbox{E}\kern-.125emX}}
\begin{document}


\title{Co-Optimization of Analog Kolmogorov-Arnold Networks for Low-Power Function Approximation in Flexible Electronics}

\author{Paula Carolina Lozano Duarte,  Georgios Zervakis, \IEEEmembership{Member, IEEE}, Mehdi Tahoori, \IEEEmembership{Fellow, IEEE}, \\and
Sani Nassif, \IEEEmembership{Life Fellow, IEEE}
\thanks{This work has been supported by the European Research Council (ERC) (Grant No. 101052764). 
(Corresponding author: Paula Carolina Lozano Duarte).}
\thanks{P. C. Lozano Duarte and M. Tahoori are with the Dept. of Computer Science, Karlsruhe Institute of Technology, Karlsruhe 76131, Germany. (e-mail: paula.duarte@kit.edu; mehdi.tahoori@kit.edu)}
\thanks{G. Zervakis is with the School of Electrical and Computer Engineering, National Technical University of Athens, Greece. (e-mail: zervakis@ece.ntua.gr)}
\thanks{S. Nassif is with Radyalis LLC, USA (e-mail: srn@radyalis.com)}
}
\maketitle

\input{Section/0_Abstract}


\bstctlcite{IEEEexample:BSTcontrol}
\thispagestyle{firstpage}
\input{Section/1_Introduction}
\input{Section/2_Background}
\input{Section/3_Methodology}
\input{Section/4_Evaluation}
\input{Section/5_Conclusion}

\bibliographystyle{IEEEtran}
\bibliography{references}

\begin{IEEEbiography}[{\includegraphics[width=1in,height=1.25in,clip,keepaspectratio]{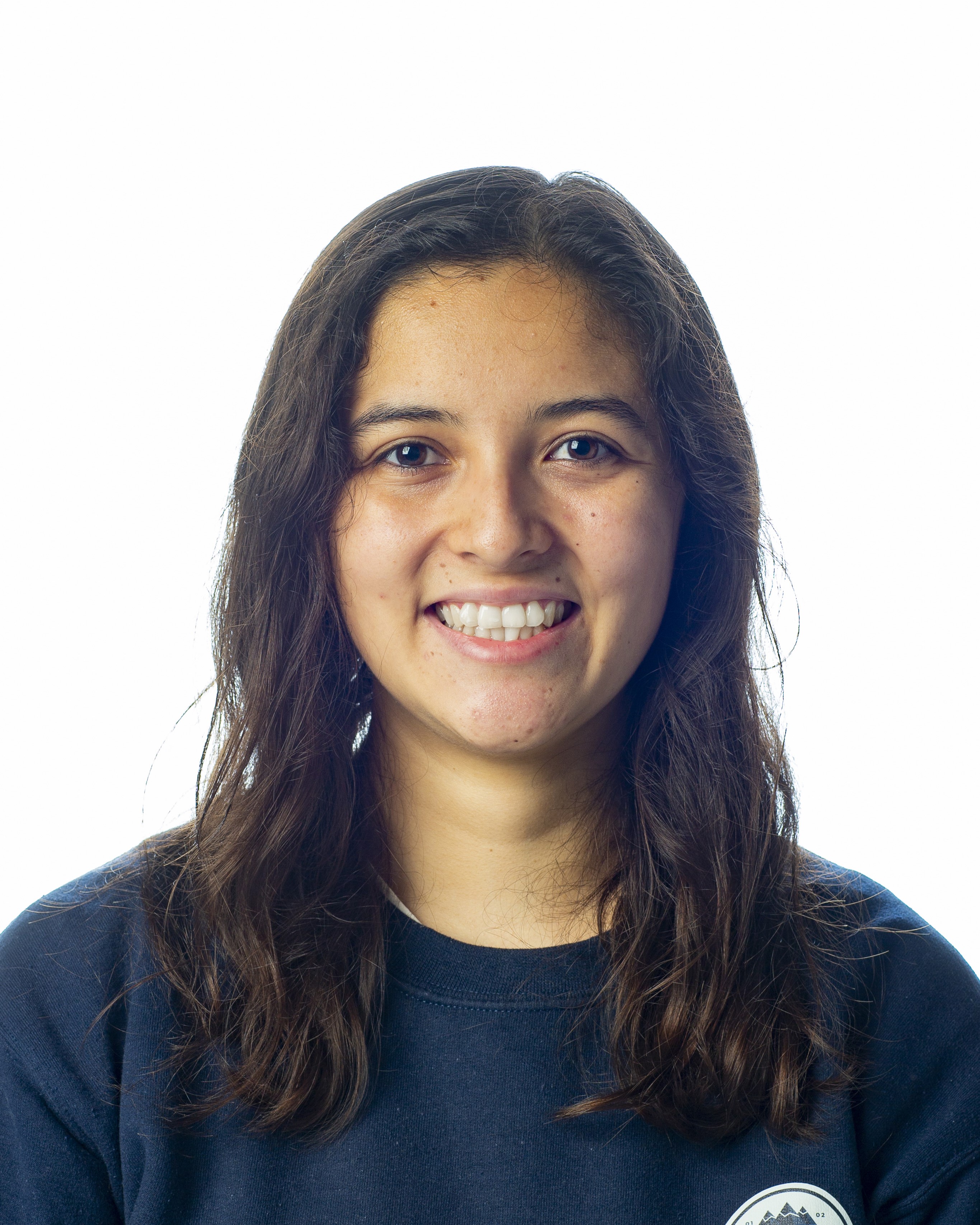}}]{Paula Carolina Lozano Duarte} is a Ph.D. student at the Chair of Dependable Nano-Computing at the Karlsruhe Institute of Technology, Germany. She received her Bachelor's degree (2021) and Master's degree (2023) in Telecommunications Engineering from the Public University of Navarra, Spain. Her main research interests include printed and flexible electronics, low-power designs, and analog computing.
\end{IEEEbiography}

\begin{IEEEbiography}[{\includegraphics[width=1in,height=1.25in,clip,keepaspectratio]{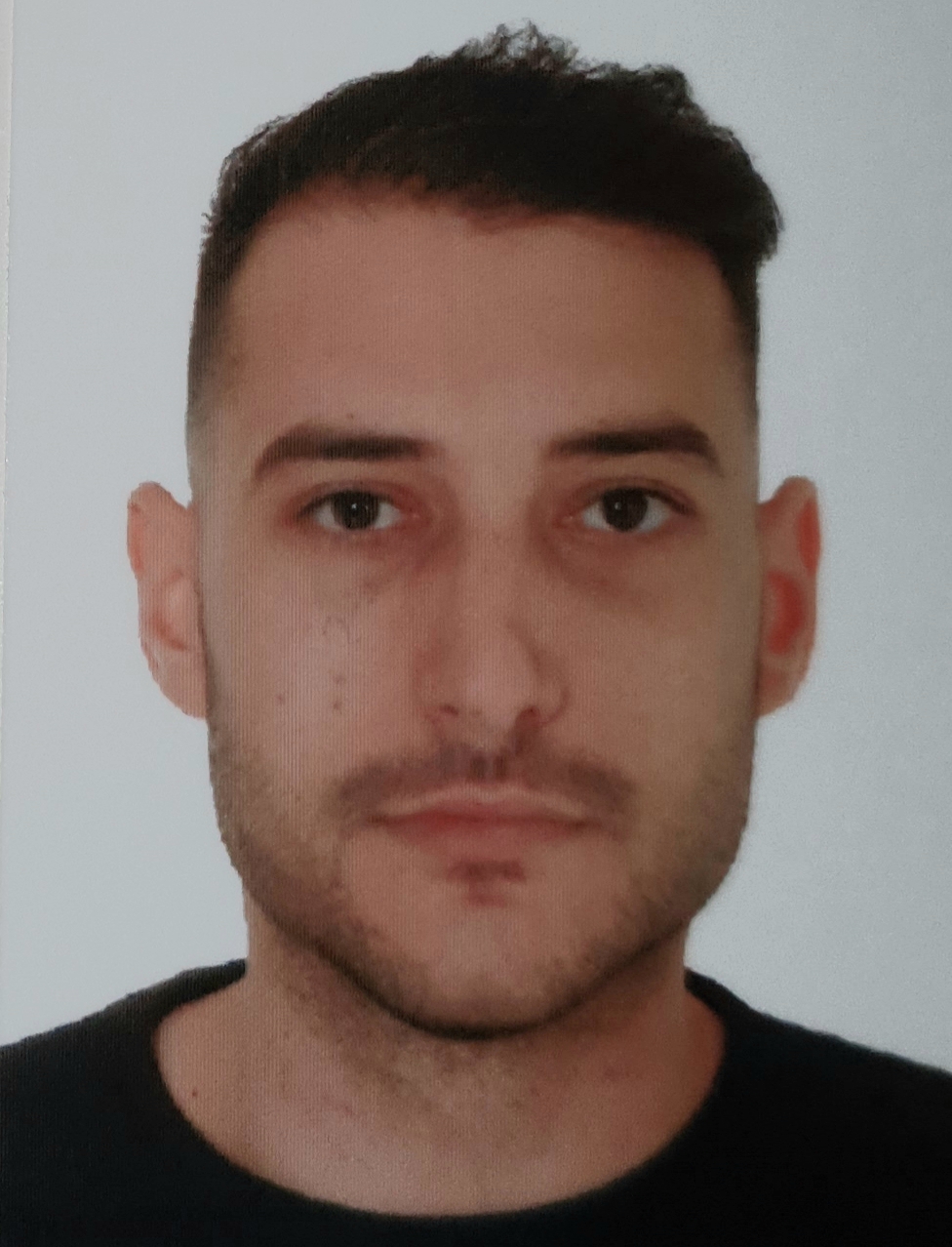}}] {Georgios Zervakis} is an Assistant Professor at the School of Electrical and Computer Engineering, National Technical University of Athens (NTUA), Greece. From 2019 to 2022, he was a Research Group Leader at the Chair for Embedded Systems (CES), Karlsruhe Institute of Technology (KIT), Germany, and from 2022 to 2025 an Assistant Professor at the Department of Computer Engineering \& Informatics, University of Patras, Greece. He received the Diploma and Ph.D. degrees from the School of ECE, NTUA in 2012 and 2018, respectively. Dr. Zervakis is an Associate Editor of IEEE TCAD and serves on the technical program committee of several major design conferences, including DAC, DATE, and ICCAD. He received a best paper nomination at DATE 2022. His research interests include digital and low-power design, electronic design automation, approximate computing, emerging technologies, printed and flexible electronics, and machine learning accelerators.
\end{IEEEbiography}

\begin{IEEEbiography}[{\includegraphics[width=1in,height=1.25in,clip,keepaspectratio]{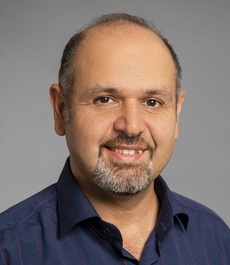}}]{Mehdi Tahoori} received the B.S. degree in computer engineering from the Sharif University of Technology, Tehran, Iran, in 2000, and
the M.S. and Ph.D. degrees in electrical engineering from Stanford University, Stanford, CA, USA, in 2002 and 2003, respectively. He is currently a Full
Professor at the Karlsruhe Institute of Technology, Karlsruhe, Germany, as well as the guest professor at imec, Leuven, Belgium. He was a recipient of the National Science Foundation Early Faculty Development (CAREER) Award. He has received a number of best paper awards at various conferences and journals, including ICCAD, FPL, ACM TODAES, and IEEE Transactions on Very Large Scale Integration (TVLSI) Systems. He was also a recipient of the European Research Council (ERC) Advanced Grant.
\end{IEEEbiography}

\begin{IEEEbiography}[{\includegraphics[width=1in,height=1.25in,clip,keepaspectratio]{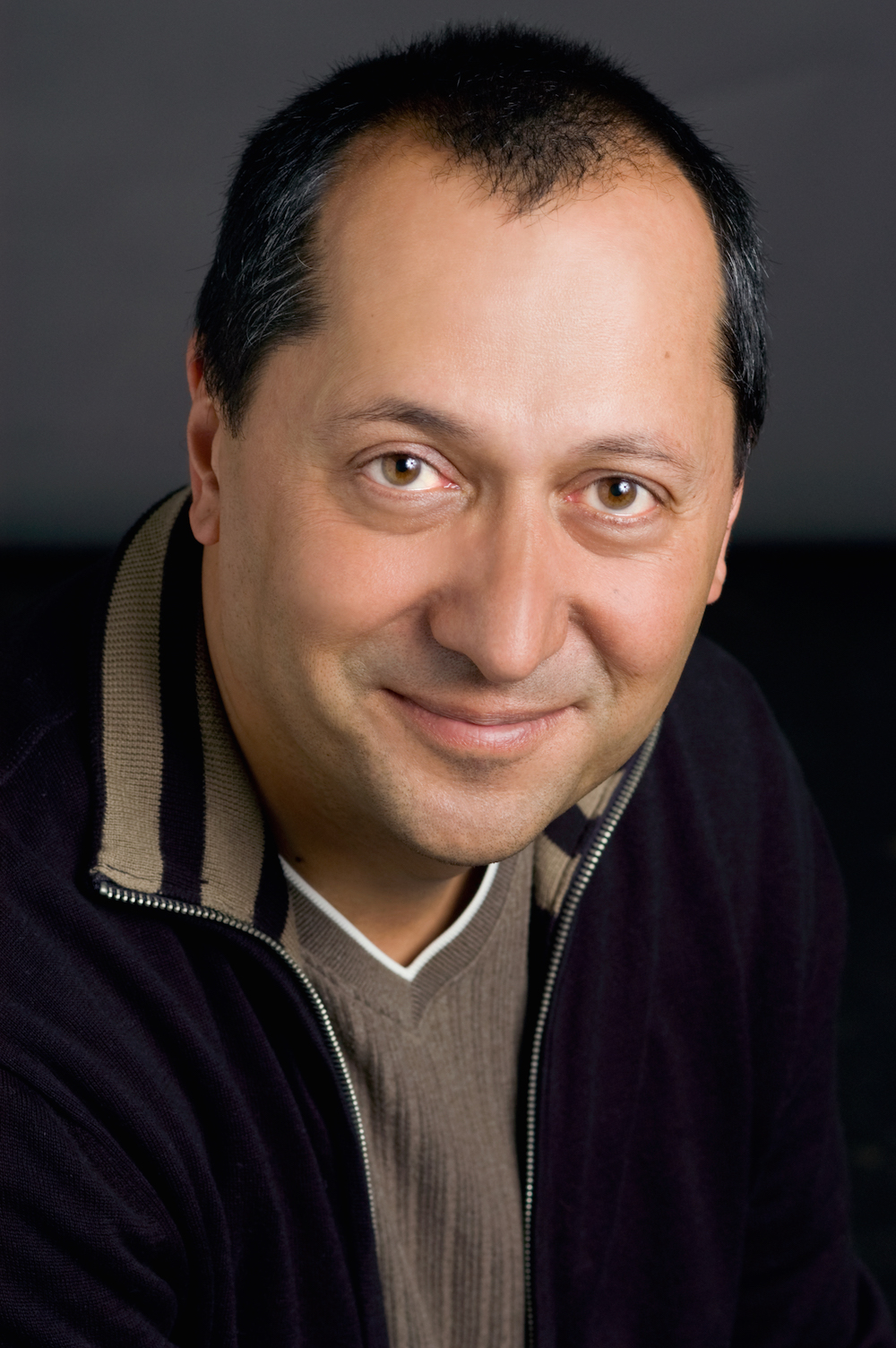}}] {Sani Nassif} Sani received his Bachelors degree from the American University of Beirut in 1980, and his Masters and PhD degrees from Carnegie-Mellon University in 1981 and 1986 respectively. He worked at Bell Laboratories until 1996, then joined the IBM Austin Research Laboratory where he had various technical and management positions. In 2014 he formed the Radyalis company which is focused on applying electronic design automation techniques to the area of Cancer Radiation Therapy. He has authored one book and numerous conference and journal publications, received many Best Paper awards (IEEE Trans. CAD, ICCAD, DAC, ISQED, ICCD, SEMICON) and given Keynote and Plenary presentations at DATE, ESSCIRC, ASPDAC, SISPAD, SEMICON, PATMOS, VLSI-SOC, GLVLSI, and ASAP. He is a Fellow of the IEEE (2008), was president of IEEE-CEDA, a member of the IBM Academy of Technology, a member of the ACM and AAAS, and an IBM Master Inventor with more than 75 patents.
\end{IEEEbiography}

\end{document}

%% file: Section/0_Abstract.tex
\begin{abstract}
\green{Wearable devices and Internet of Things (IoT) sensors require on-sensor processing of biosignals and environmental data, including computationally demanding operations such as nonlinear activation functions for neural network inference, sensor calibration curves \orange{to map} raw readings to physical units, and signal preprocessing functions like logarithmic compression and power operations for feature extraction. These functions exhibit significant complexity, often involving transcendental operations and multivariate dependencies that are costly to implement digitally.}
Analog function approximation provides a power-efficient alternative by performing these computations in the analog domain, thereby reducing the energy overhead associated with analog-to-digital conversion and subsequent digital processing.
\green{Flexible Electronics (FE) present a particularly attractive platform for wearable applications due to mechanical flexibility and low-cost fabrication, but impose strict constraints on circuit density and power consumption, making efficient analog implementations critical but challenging.}
This work introduces Analog Kolmogorov-Arnold Networks (AKANs), developed via hardware–software co-optimization, to approximate \green{these complex multivariate} functions accurately under hardware imperfections. 
Our method incorporates circuit-level error modeling during training and applies pruning at both software and hardware levels to reduce area and power. 
Validation across multiple benchmarks demonstrates that \green{our proposed pruning methodology} not only reduces hardware cost but can also improve approximation accuracy by regularizing spline parameters. Results show up to 55\% area and 50\% power savings, with average reductions of nearly 30\% across datasets, highlighting AKANs as a robust and generalizable framework for low-power analog function approximation in FE.
\end{abstract}

%% file: Section/1_Introduction.tex
\section{Introduction}\label{sec:intro}

\black{Emerging applications such as wearable devices~\cite{junaid2022healthcare}, healthcare monitoring~\cite{liu2024wearable}, and compact Internet of Things (IoT) sensors~\cite{shumba2022iot} require on-sensor processing to address the stringent constraints on data transmission, power efficiency, and real-time responsiveness~\cite{gupta2020radiology}. These applications often operate in resource-constrained environments, where transmitting raw sensor data to digital processing units is too power-hungry and inefficient.} Analog pre-processing of sensor data offers a promising alternative by performing the first layer of computation before digitization, thus reducing power consumption, latency, and area requirements~\cite{analog2022power,fierce2020analog,techbriefs2020sensing}.

Flexible Electronics (FE) provide a promising platform for achieving power-efficient sensor processing, offering advantages such as mechanical flexibility, lightweight design, and cost-effective fabrication~\cite{Baruah:FabricationFE2023, Jeong:igzoperformance}. 
However, implementing high-performance computational models in FE presents significant challenges due to limitations in circuit density and power consumption~\cite{Arokia:feAdvantages2012, Lozano:aspdac25:BinCoDesign}.

\black{Analog function approximation plays a crucial role in addressing these challenges, enabling efficient computation on resource-constrained hardware.} 
\orange{In wearable and IoT systems, function approximation is routinely required for tasks such as nonlinear activation functions in neural networks, sensor calibration, and analog signal conditioning. Typical operations include sigmoids, tanh, and other smooth activations~\cite{yang2025efficient}, polynomial mappings for sensor linearization in flexible analog front ends~\cite{Zikang:analogfrontendsflexible, Shi2023IntegratedAF}, and compression or power-law transformations for feature extraction~\cite{Chen:analogoptimization, sadasivuni2022insensor}. 
These functions are increasingly implemented directly in analog hardware to reduce digital workload in near-sensor or in-sensor computing architectures~\cite{li2015rram, agarwal2016energy}.}
\blue{Analog implementations remain vulnerable to noise, process variation, and device mismatch, which can significantly degrade approximation accuracy if hardware effects are not explicitly considered during model design~\cite{lozano2026faulttolerantdesignigzobased}. 
These constraints motivate the need for co-design methodologies that jointly optimize algorithmic accuracy and hardware efficiency while maintaining robustness to analog non-idealities within the limitations of FE technologies.}

\black{Several state-of-the-art approaches have been proposed to address these trade-offs in conventional and emerging technologies: RRAM-based computing-in-memory has shown promise for enabling efficient matrix operations, but device variability and endurance remain significant limitations~\cite{yang2025efficient, li2015rram, agarwal2016energy, ielmini2025resistive}. Polynomial-based techniques, particularly Chebyshev approximations, provide predictable accuracy with low hardware cost, though their energy demand increases with function complexity~\cite{us8832168b2, moshfe2025programmable, yu2023approximate}. Meanwhile, analog neural networks with linear kernels offer energy-efficient mappings of nonlinear functions~\cite{gao2024kirchhoffnet, sadasivuni2022insensor}.}

Kolmogorov-Arnold Networks (KANs) provide a structured and automated approach to function approximation, eliminating the need for manually designed architectures and enabling efficient representation of complex functions~\cite{Liu:KAN2024,Sidhart:fuctionapproxKAN2024}. 
By leveraging a decomposition of multivariate functions into a sum of univariate functions, KANs reduce the complexity of function approximation, making them a compelling alternative to traditional polynomial or piecewise-linear methods. However, direct hardware implementations of KANs often incur significant area and power overhead, limiting their applicability in FE platforms~\cite{VanDuy:KANCodesign:2024, Huang:KANAcceletaror:2024}.

\blue{To address these limitations, this work proposes an Analog Kolmogorov-Arnold Network (AKAN) designed through a hardware--software co-optimization methodology that explicitly accounts for circuit-level constraints during model training and pruning. The proposed framework integrates circuit simulation, error modeling, and parameter pruning into a unified design loop, enabling systematic exploration of accuracy--cost trade-offs in FE implementations.
The central novelty of this work rests on three key contributions:

\begin{enumerate}
    \item \textbf{Technology-aware analog error modeling for FE.} We derive continuous, input-dependent error functions directly from transient SPICE simulations of unipolar IGZO TFT circuits, capturing systematic hardware non-idealities in a commercially relevant FE process.

    \item \textbf{Circuit-block-level coefficient pruning co-designed with KAN training.} Our pruning operates at the level of individual polynomial coefficients ($k_0$, $k_1$, $k_2$) of each quadratic spline, where each removed coefficient corresponds to the physical removal of a circuit block (multiplier, squarer, or adder) in the IGZO implementation.

    \item \textbf{Application-driven validation on representative sensor datasets.} The complete co-design methodology is evaluated across four datasets—PPG-DaLiA, ECG5000, Individual Household Electric Power Consumption, and Iris—covering biomedical monitoring and sensing scenarios relevant to FE deployment.
\end{enumerate}
}

\blue{The proposed methodology differs fundamentally from prior digital and software-based KAN implementations. Existing approaches typically assume discrete digital hardware models or apply statistical error distributions independent of circuit behavior. In contrast, our framework derives a continuous error function $\hat{\epsilon}_{\text{hardware}}(x)$ directly from transient circuit simulations and updates this model for each pruning configuration. This closed-loop interaction between circuit topology and training objective ensures that pruning decisions are validated through measurable hardware metrics, including area and power consumption, rather than purely algorithmic criteria.}

\blue{This manuscript represents a substantial extension of our preliminary conference paper~\cite{duarte2025functionapproximationusinganalog}. The prior work introduced the basic analog building blocks and a fixed-offset error model for a single spline configuration. The present work expands that foundation through:

\begin{itemize}
    \item a systematic hardware characterization covering 1,000 SPICE simulation runs across the full coefficient space ($k_0$, $k_1$, $k_2$),
    \item a continuous, input-dependent error model $\hat{\epsilon}_{\text{hardware}}(x)$ derived from simulation data and embedded into the KAN training process,
    \item a coefficient-level pruning methodology evaluating all 20 distinct spline configurations with direct mapping to circuit-block removal, and
    \item validation across multiple sensor datasets demonstrating substantial reductions in hardware cost while maintaining approximation accuracy.
\end{itemize}
}

\textit{The combination of pruning and hardware-aware optimization ensures that AKAN implementations remain feasible for resource-constrained environments while maintaining accuracy comparable to digital counterparts.}
\textcolor{black}{The proposed coefficient-level pruning reduces the area and power consumption of the splines by up to 55\% and 50\%, respectively, relative to the unpruned baseline. These figures represent block-level improvements; the system-level impact depends on the fraction of the complete sensing pipeline occupied by this analog building block, see \autoref{sec:results}.}

The rest of the paper is structured as follows: ~\autoref{sec:background} provides the necessary background about FE, analog function approximation, and KAN-based implementations. \autoref{sec:methodology} details the proposed AKAN design methodology, including hardware characterization, error modeling, pruning strategies, and training procedures. \autoref{sec:results} presents the experimental evaluation of pruning configurations and their impact on accuracy, area, and power consumption. Finally, ~\autoref{sec:conclusion} concludes the paper.

%% file: Section/2_Background.tex
\section{Background}\label{sec:background}

\subsection{Flexible Electronics}

\begin{figure*}
    \centering
    \includegraphics[width=1\textwidth]{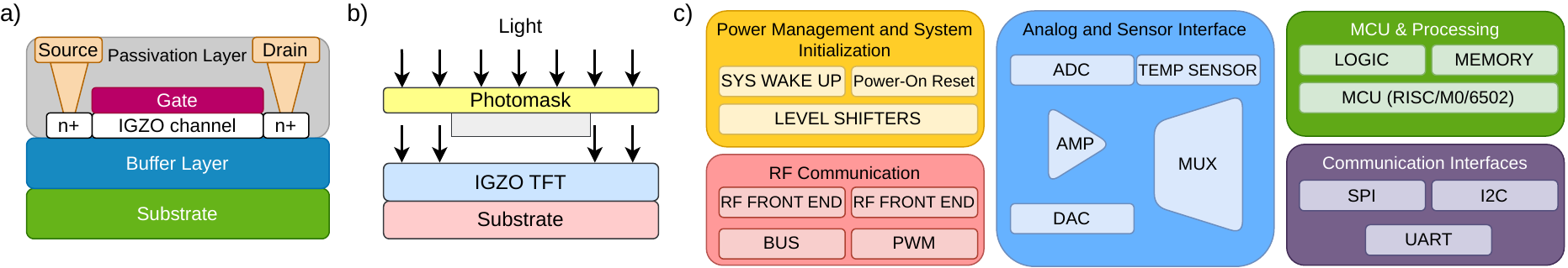}\vspace{-2ex}
    \caption{a) Cross-section of a unipolar IGZO TFT showing key layers. b) Process schematic illustrating photolithographic patterning on flexible substrates, typically involving deposition of IGZO semiconductors and metallic interconnects c) PragmatIC's FlexIC design capabilities, integrating power management, analog/sensor interfaces, processors, and communication modules into flexible ICs~\cite{PragmatICFlexICs}.}\label{fig:FE_background}\vspace{-2ex}
\end{figure*}

FE are systems designed to bend and conform to non-rigid surfaces while retaining functionality. Fabricated on substrates like polyimide, plastic, or metal foil, they support applications in wearable health monitors, adaptive sensors, and flexible displays~\cite{Heng:AM2022:FlexHumanMachInterfaces,Bonnassieux:FPE:2021:FlexiblePrintedRoadmap}. Their adaptability makes FE ideal for next-generation lightweight and versatile electronic systems.

Indium Gallium Zinc Oxide (IGZO) is central to FE, enabling thin-film transistors (TFTs) with high mobility and low leakage. IGZO outperforms amorphous silicon but lacks P-type transistors, leading to unipolar circuit designs and requiring techniques like dynamic logic and pseudo-CMOS to achieve full functionality~\cite{Zhu:IGZO2021,Pan:IGZOTFT2024}, see Fig.~\ref{fig:FE_background}a.

Adaptations of semiconductor processes to flexible substrates reduce costs and complexity. PragmatIC Semiconductor's FlexIC platform avoids ion implantation and high-temperature steps, enabling cost-effective mass production with critical dimensions around 600 nm~\cite{EuropracticeFlexibleElectronics}. 
FlexICs can integrate MCUs~\cite{Ozer:2024:RISCmicro}, sensors~\cite{costa2023natively}, communication protocols (NFC, SPI, I2C, UART), and analog circuits (ADCs, DACs)~\cite{lozano2026faulttolerantdesignigzobased, Alkhalil:BioCAS:2022:FlexibleSAR}, signal conditioning circuits~\cite{Zulqarnain:hpfigzo,Garripoli:analogfrontendigzo}, enabling tailored solutions for application-specific requirements, see Fig.~\ref{fig:FE_background}c.
These blocks support ultra-low-power signal processing for real-time feature extraction in wearables and IoT sensors~\cite{Gao:FlexibleWearableSensing,Baruah:FabricationFE2023}.

While not a replacement for silicon, FE complements it in domains demanding stretchable, disposable, or low-cost solutions like smart packaging, wearables, and interactive surfaces~\cite{Luo:smartpack,lee:flexiblepatch}. This expands electronic applications into areas inaccessible to rigid ICs.

\subsection{Analog Function Approximation}

Analog function approximation is a fundamental aspect of analog computing, enabling the representation and processing of mathematical functions using continuous electrical signals~\cite{wright2021deep}. 
Historically, analog function approximation has been pivotal in applications requiring real-time processing and low-power consumption, such as signal processing and control systems~\cite{analog2022power,techbriefs2020sensing}.

Recent advancements have revitalized interest in analog computing for function approximation. One of these is the development is the integration of nonlinear activation functions within analog resistive crossbar arrays~\cite{yang2025efficient}. 
This innovation facilitates efficient implementation of complex mathematical operations directly in hardware, enhancing the performance of analog neural networks. 
By embedding nonlinear functions into the hardware architecture, these systems can achieve faster computation speeds and reduced energy consumption compared to their digital counterparts. 

\black{Complementary studies have shown that resistive memory devices (RRAM) can be exploited not only for storage but also for analog approximate computing, reducing hardware overhead while supporting neural inference tasks~\cite{li2015rram, agarwal2016energy, ielmini2025resistive}. Such RRAM-based implementations highlight the dual role of memory and computation in enabling compact, energy-efficient analog architectures.}

Another significant contribution is the use of analog in-memory computing (AIMC) for kernel approximation in machine learning algorithms~\cite{buchel2024kernel}. This approach leverages the inherent parallelism and low latency of AIMC architectures to perform complex computations more efficiently. By integrating computation and memory storage, AIMC reduces data movement and accelerates processing times, offering a promising solution for energy-efficient machine learning applications. 
\black{In parallel, polynomial-based analog approximations, such as Chebyshev expansions, have been investigated as an alternative to memory-centric approaches. These methods provide predictable error bounds and can be implemented in hardware-friendly architectures, although energy cost scales with polynomial degree~\cite{us8832168b2, moshfe2025programmable, yu2023approximate, wang2018timedomainanalogweightedsumcalculation}.}

Additionally, the development of compact analog computing systems, such as those utilizing ultrasonic wave interactions, has expanded the capabilities of analog function approximation~\cite{Uy2023:diffeqinanalogcomp}. These systems can solve a variety of linear and nonlinear integro-differential equations, demonstrating the versatility of analog approaches in modeling complex physical phenomena. 
\black{Analog neural network architectures have also been investigated in the form of Kirchhoff-based networks and in-sensor computing schemes, enabling energy-efficient mappings of nonlinear functions~\cite{gao2024kirchhoffnet, sadasivuni2022insensor}.}

\subsection{Kolmogorov-Arnold Networks: Theory and Architecture}

\input{FigTab/kan_bg}

KANs are a class of artificial neural networks that leverage the principles of Kolmogorov's superposition theorem~\cite{theorem}. This theorem states that any continuous function of several variables can be represented as a finite superposition of continuous functions of a single variable. Unlike conventional neural networks that rely on layered compositions of activation functions, KANs exploit this theorem to decompose complex functions into structured, hierarchical representations, enabling them to model nonlinear relationships effectively. 
In Fig.~\ref{fig:KAN_background}, we show the architecture of Multi-Layer Perceptrons and KANs, contrasting their respective layer structures and activation function implementations.

The architecture of KANs typically consists of multiple layers, where each layer performs a specific transformation of the input data. The first layer applies basis functions to the inputs, generating a set of intermediate values. These intermediate values are then processed through subsequent layers to refine the approximation. By optimizing the parameters of the network, KANs can learn to approximate a wide range of functions, making them suitable for applications in fields such as signal processing and control systems~\cite{Liu:KAN2024,Sidhart:fuctionapproxKAN2024}. The hierarchical nature of these networks allows them to efficiently capture fine-grained variations in the data while maintaining computational efficiency.

A key advantage of KANs is their ability to achieve high approximation accuracy with relatively few parameters, reducing memory and computational overhead compared to traditional deep neural networks. Furthermore, their inherent structure enhances robustness to noise and uncertainties in the input data, making them suitable for low-power, real-time applications where computational efficiency is crucial~\cite{Sidhart:fuctionapproxKAN2024,duarte2025functionapproximationusinganalog}.
As illustrated in Fig.~\ref{fig:KAN_background}, the architecture of a generic KAN consists of multiple spline-based approximators interconnected through Multiply-Accumulate (MAC) operations.

\textcolor{black}{Several recent works have explored hardware implementations of KANs beyond the digital CMOS domain. 
Wen et al.~\cite{wen2026kan_cim} propose a computing-in-memory (CIM) architecture based on tunable Gaussian-like memory cells composed of anti-ambipolar transistors and memristors, enabling analog in-memory evaluation of KAN activations with significant energy efficiency gains over GPU implementations. Sudarshan et al.~\cite{sudarshan2026kacim} introduce KA-CIM, a CIM accelerator that computes arbitrary nonlinear functions using a piecewise-linear approximation scheme, demonstrating that KAN inference can be efficiently mapped onto memory-centric architectures. Taglietti et al.~\cite{taglietti2026kan_physical} further demonstrate physical KAN implementations in silicon-on-insulator devices, showing that learning nonlinear device dynamics, rather than linear weights, can improve task performance per physical resource.}
\textcolor{black}{Collectively, these works establish analog and physical KAN implementations as an active research direction. However, none of them target FE or IGZO-based wearable platforms, leaving a gap in ultra-low-power, mechanically constrained implementations.}

\textcolor{black}{In parallel, prior work on KAN compression and acceleration has primarily focused on digital and algorithmic approaches. Huang et al.~\cite{Huang:KANAcceletaror:2024} investigate sparsity-aware mapping and quantization-aware training for efficient KAN inference on resource-constrained CMOS hardware, targeting lightweight edge deployment.}
\textcolor{black}{ShapKAN~\cite{Fan:ShapKAN:2025} proposes a software-only pruning framework for KANs based on shift-invariant Shapley-value attribution. In this approach, the importance of each KAN node is estimated using a game-theoretic contribution score that is designed to be invariant to permutation and scaling effects in KAN representations. Nodes with low attributed importance are removed to achieve model compression.}
\textcolor{black}{While both ShapKAN and the present work aim at reducing KAN complexity, they differ fundamentally across multiple dimensions. First, the pruning criterion in ShapKAN is based on global attribution scores derived from Shapley values, whereas the present work uses SPICE-validated numerical error metrics (NMPE) tied to circuit-level behavior. Second, ShapKAN operates at the node level, removing entire functional units, while the present work performs coefficient-level pruning within spline functions, enabling finer-grained control aligned with physical circuit decomposition.}
\textcolor{black}{Third, ShapKAN is purely algorithmic and hardware-agnostic, whereas the present approach is explicitly hardware-aware, incorporating closed-loop circuit simulation into the pruning process. Finally, ShapKAN is evaluated on digital inference accuracy metrics, while the present work targets FE wearable sensing applications where performance is jointly constrained by accuracy, power, and area in an IGZO-based analog implementation.}

\subsection{Analog Implementation of KANs in FE}

\input{FigTab/implementation}

The motivation for implementing KANs in the analog domain when using FE, as opposed to digital methods, is primarily driven by the constraints and demands of FE. Digital implementations of function approximations can become cumbersome and costly, particularly in applications such as wearable devices and IoT sensors that require real-time processing with minimal energy expenditure.

Analog Building Blocks (ABBs) are specifically designed to perform fundamental functions such as inversion, subtraction, addition, multiplication, and squaring. Each operation is optimized to enhance functionality while addressing constraints inherent in FE, establishing a solid foundation for the KAN implementation~\cite{duarte2025functionapproximationusinganalog}. 

The ABBs, illustrated in Fig.~\ref{fig:implementation} a) to e), provide the basic arithmetic operations necessary for constructing the function approximators. By combining these blocks efficiently, more complex functions such as quadratic splines can be realized.

\blue{The hardware implementation realizes the following quadratic polynomial representation:}

\begin{equation}
\mathbf{B}(x) = P_0 + (P_1 - P_0)2x + (P_0 - 2P_1 + P_2)x^2
\label{eq:quadratic_formula}
\end{equation}

which enables the construction of splines with fewer blocks compared to alternative approaches, such as the Bezier second-order representation~\cite{Liu:KAN2024}. Specifically, the implementation requires five blocks: two multiplications, one squaring, and two additions. Each sub-block of the spline is designed to ensure that input ranges are compatible with overall system requirements, maintaining functionality within specified limits. The complete analog spline implementation is shown in Fig.~\ref{fig:implementation} f).

Additionally, the KAN is constructed by applying MAC operations, facilitated by strategically connecting resistors to the splines. This design allows for the adjustment of resistor values to achieve desired weights for each spline, ensuring flexibility in the KAN's architecture. This modular approach not only simplifies the retraining process---allowing updates to input values without a complete redesign---but also lays the groundwork for future optimizations. The architecture of the analog KAN is depicted in Fig.~\ref{fig:implementation} g).

To further enhance performance and minimize error, various training and optimization techniques can be applied to AKANs, including hardware-aware pruning strategies and adaptive resistor tuning. These techniques enable a balance between accuracy and efficiency, making them particularly well-suited for energy-constrained environments such as flexible and wearable electronics. 

Nevertheless, this design can be enhanced through hardware and software co-optimization. Our work aims to further reduce area and power by conducting a comprehensive exploration of pruning methods for the splines and their impact on the overall accuracy of the approximation.

%% file: FigTab/kan_bg.tex
\begin{figure*}[t]
\centering
\includegraphics[width=0.85\linewidth]{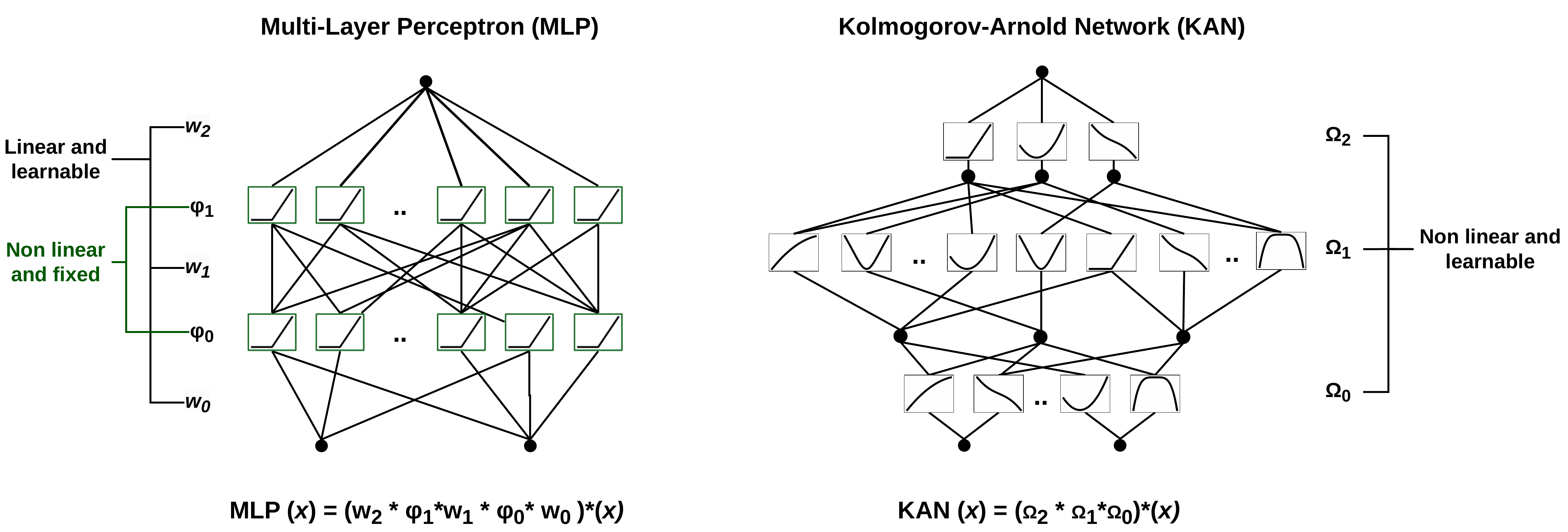}
\caption{Comparison of the MLP and KAN architectures, highlighting the differences in their layer structures and activation function implementations.}\label{fig:KAN_background}\vspace{-2ex}
\end{figure*}

%% file: FigTab/implementation.tex
\begin{figure}[t]
\centering
\includegraphics[width=1\linewidth]{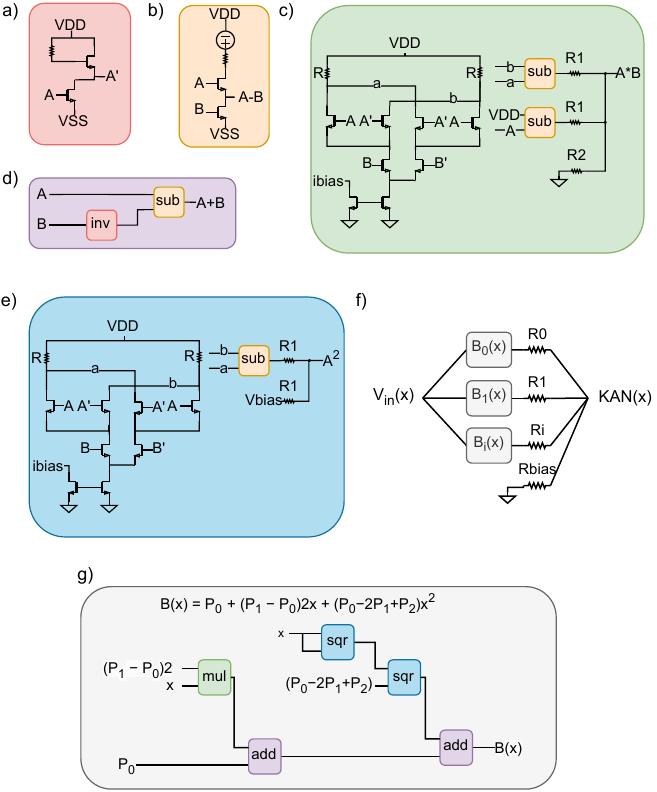}\vspace{-2ex}
\caption{Schematic representation of the Analog Building Blocks: a) Inversion, b) Subtraction, c) Multiplication, d) Addition, and e) Squaring. Architectural overview of f) an AKAN utilizing \( i \) splines. The KAN is constructed using MACs, where resistors connected in series with each spline enable weighted accumulation. The resistor values determine the contribution of each spline, allowing flexibility in weight assignment. g) an Analog Spline.}\vspace{-2ex}

\label{fig:implementation}\vspace{-2ex}
\end{figure}

%% file: Section/3_Methodology.tex
\section{Proposed Hardware-Software Co-Design of the AKAN}\label{sec:methodology}

\input{FigTab/flowchart}
\green{This section presents the hardware-software co-design methodology for pruning-aware AKAN implementations. The approach consists of four main stages: (i) hardware architecture design, (ii) circuit-level error characterization and cost analysis, (iii) pruning-aware training strategy, and (iv) validation across application-relevant benchmarks.}
The proposed approach builds upon existing architectures that use ABBs for efficient spline approximation.
\green{Our key contribution is a systematic pruning methodology that operates at the spline parameter level, enabling joint optimization of hardware cost (area and power) and approximation accuracy. Unlike conventional pruning strategies that remove entire splines or neurons, we selectively \orange{set} individual polynomial coefficients ($k_0$, $k_1$, or $k_2$) \orange{to zero} based on their contribution to the approximation accuracy.}

\blue{Three features distinguish this framework from generic hardware-in-the-loop training and from the CIM-oriented KAN work of~\cite{Huang:KANAcceletaror:2024}. First, the error function $\hat{\epsilon}_{\text{hardware}}(x)$ is derived from transient SPICE simulations of unipolar IGZO TFT circuits, whose operating regime (subthreshold leakage, low $V_{\text{TH}}$ uniformity, absence of P-type devices) produces input-dependent error profiles qualitatively different from SRAM- or RRAM-based CIM arrays. Second, pruning operates at the coefficient level within each spline rather than at the node level, and each pruned coefficient corresponds to the physical removal of a discrete circuit block---creating transistor-level area and power savings. Third, unlike~\cite{Huang:KANAcceletaror:2024} where the error model is applied after architecture design is fixed, our error model is updated per pruning configuration, creating a closed loop between circuit topology and training objective.}

An additional distinguishing feature of our approach is the incorporation of a simulation-based error model, which captures the error variations across different input signals. The error function derived from these simulations is integrated into the KAN training process to account for systematic errors introduced by pruning. 
\blue{A layered overview of the proposed hardware-aware training and pruning workflow for AKAN-based function approximation is shown in Fig.~\ref{fig:flowchart}.}
\green{Finally, we validate the framework on different sensor datasets, demonstrating its robustness for FE-based preprocessing applications.}

\subsection{AKAN hardware architecture}

\input{FigTab/pruning}

For the hardware design of our AKAN, we use as a baseline an existing architecture that utilizes ABBs for function approximation\green{~\cite{duarte2025functionapproximationusinganalog}}. 
\green{While the baseline circuit topology remains unchanged, our contribution lies in: (1) systematic characterization of the parameter space through \orange{a large number of circuit} simulation runs to establish coefficient-to-error mappings; (2) development of a pruning methodology that selectively removes polynomial terms rather than entire splines; and (3) integration of circuit-level error models into KAN training for hardware-aware optimization.} 
After replicating the baseline design, we perform parametric sweeps across all spline coefficients to characterize circuit behavior.
Using~\autoref{eq:quadratic_formula} as the foundation for each spline, we redefine the constant terms to simplify the equation. This approach allows us to express the spline as a second-degree polynomial:

\begin{equation}
k_0 + k_1 x + k_2 x^2,
\end{equation}
where
\begin{align}
k_0 &= P_0, \\
k_1 &= (P_1 - P_0) \times 2 \\
k_2 &= P_0 - 2P_1 + P_2.
\end{align}
\green{These transformations map the piecewise control points $P_0, P_1, P_2$ to polynomial coefficients that directly correspond to circuit bias currents and transistor sizing ratios in the ABB implementation.}

\blue{The central design objective of this work is to identify pruning configurations that reduce hardware cost while preserving approximation fidelity within predefined accuracy bounds.}
To this end, we employ a hardware-software co-design approach, with pruning serving as one of the key parameters for optimizing the hardware. 
\orange{
\blue{Pruning is evaluated at two complementary levels:}
\begin{itemize}
    \item \textbf{Spline level:} individual coefficients ($k_0$, $k_1$, or $k_2$) are removed, and the resulting circuit simplification is evaluated in terms of area and power.
    \item \textbf{KAN level:} the impact of these local coefficient removals on overall approximation accuracy is assessed across the complete input space.
\end{itemize}

By evaluating both hardware implications and system-level performance for each configuration, this hardware-aware methodology identifies pruning strategies that reduce circuit complexity while maintaining acceptable approximation quality.
}

Each parameter ($k_0, k_1, k_2$) has a different impact on the hardware design. Removing one parameter may lead to a reduction in hardware area and power, but it also introduces a trade-off in the approximation accuracy of the spline. The structural modifications in the spline design, depending on the parameter removed, are illustrated in Fig.~\ref{fig:pruning}, showing the varying degrees of complexity for each configuration. \green{The impact on area, power, and approximation error is quantified through circuit-level simulations to establish Pareto-optimal pruning configurations.}

Since the input range for $k_0, k_1, k_2$, and $V_{in}$ is constrained between -0.5V and 0.5V (equivalent to $V_{dd}/2$), all simulations adhere to this limitation.

\subsection{Error Modeling and Hardware Cost Evaluation}

In this phase, we aim to quantify the error and hardware cost of the realization of various splines. The parameters \(k_0\), \(k_1\), and \(k_2\) are systematically swept across 10 discrete values within the range of -0.5V to 0.5V.
For each set of parameters, simulations are conducted using the Cadence Spectre simulator, which evaluates the hardware cost in terms of area and power consumption. \green{This characterization establishes the relationship between coefficient values and both approximation fidelity and hardware efficiency.}

The simulations involve 1,000 different runs, each with a ramp input signal (\(V_{in}\)) spanning -0.5V to 0.5V, sampled at 0.02\green{V intervals over a 10ms sweep duration}, resulting in 500 distinct values per run. \green{Transient simulations use a maximum timestep of 1$\mu$s to ensure accurate capture of circuit dynamics. The circuit settles within 100$\mu$s for each input step, and measurements are taken after this settling time to avoid transient effects.} For each combination of \(k_0\), \(k_1\), and \(k_2\), we compute the NMPE between the hardware-implemented spline and the ideal mathematical spline.

\blue{The simulation data are transferred from Cadence to Python as follows. For each of the 1,000 runs, the output voltage waveform $V_{\text{out}}(x)$ is exported as a CSV file containing the swept input $V_{\text{in}}$ and the simulated $V_{\text{out}}$ at 500 uniformly spaced values in $[-0.5\,\text{V}, +0.5\,\text{V}]$. The per-sample error is defined as $\epsilon_{\text{raw}}(x_i) = V_{\text{out,SPICE}}(x_i) - \mathbf{B}_{\text{ideal}}(x_i)$, where $\mathbf{B}_{\text{ideal}}(x)$ is the ideal spline polynomial. From the 1,000 runs, the 30 configurations with NMPE closest to zero are retained. Their error profiles are aggregated into a single representative curve, which is then fitted in Python using \texttt{scipy.optimize.curve\_fit} over a candidate set of polynomial, exponential, logarithmic, and rational models. The single fitted function $\hat{\epsilon}_{\text{hardware}}(x)$---not the raw tabular data---is subsequently embedded into the KAN training loop as an additive offset on the target signal.}
\blue{The same pre-characterized simulation dataset is reused across all 20 pruning configurations, allowing rapid evaluation of alternative hardware topologies without repeating circuit-level simulations.}

\green{The resulting data identifies parameter ranges where hardware realizations achieve acceptable approximation quality (NMPE near zero) at minimal area and power cost. These findings directly inform the pruning thresholds applied in subsequent stages.}

\green{The best-fit error model is selected based on the coefficient of determination ($R^2$), with a minimum requirement of $R^2 > 0.95$ to ensure accurate representation of hardware behavior. In cases where multiple models satisfy this criterion, the model with the lowest complexity (fewest parameters) is selected to reduce the risk of overfitting.}

\blue{These characterized error profiles form the quantitative basis for pruning decisions and directly parameterize the hardware-aware training stage described in the next subsection.}

\subsection{Pruning and Training Strategy for KAN Optimization}

The pruning process begins by selecting optimal values for \(k_0\), \(k_1\), and \(k_2\) based on the error and hardware cost analysis. \green{At the KAN level, we systematically evaluate all possible pruning combinations: no pruning (baseline), pruning $k_0$ only, pruning $k_1$ only, pruning $k_2$ only, and all pairwise and triple combinations.}
\blue{For each pruning configuration, the corresponding hardware error function is recomputed using the stored circuit characterization data.}

For each pruning configuration, we adjust the spline by removing one parameter (either \(k_0\), \(k_1\), or \(k_2\)) and then re-compute the error function. The error is then modeled using the same set of functions (polynomial, exponential, logarithmic, rational), and the best-fit model is selected based on the $R^2$ value. \green{This ensures that each pruning scenario has an accurate error model for subsequent training.}

\green{The pruning strategy balances two competing objectives: minimizing hardware cost (area and power) while keeping approximation error within acceptable bounds. By comparing pruned configurations to the baseline, we identify which coefficients can be safely removed with minimal accuracy degradation.}

\blue{A defining aspect of the training methodology is the integration of hardware-aware error models into the optimization objective. Instead of assuming ideal circuit behavior, the training process explicitly accounts for measured hardware deviations, enabling the network to learn parameter configurations that remain robust under realistic operating conditions.}


The training process is conducted in Python, where the error models are integrated into the KAN training framework. \orange{The training workflow consists of three main stages:

\textbf{Stage 1: Error Model Assignment.} For each pruning configuration (baseline, single-parameter pruned, or multi-parameter pruned), we assign the corresponding error model obtained from circuit simulations. This creates a library of error functions indexed by pruning state.

\textbf{Stage 2: Noisy Target Generation.} During training, we inject hardware errors into the target function by adding the error model as additive noise. For noiseless (ideal) training, this step is skipped. For noisy training (baseline or pruned), the target becomes $f_{\text{target}}(x) = f_{\text{ideal}}(x) + \epsilon_{\text{hardware}}(x)$, where $\epsilon_{\text{hardware}}(x)$ is the fitted error function from the corresponding pruning configuration.

\textbf{Stage 3: KAN Training and Evaluation.} The KAN is trained using standard backpropagation to minimize mean squared error between predicted and noisy target values. For a KAN with 3 splines, we evaluate 20 unique pruning combinations across all target functions (accounting for symmetry, where the order of pruned splines does not affect the result). Each combination corresponds to a distinct spline configuration (baseline, single-parameter pruned, or multi-parameter pruned), allowing us to map the full design space of hardware cost versus approximation accuracy. Performance is quantified using NMPE on the validation set.

This approach forces the KAN to implicitly compensate for hardware imperfections during training through adaptive weight adjustment, rather than post-hoc error correction. By comparing noiseless, baseline noisy, and pruned noisy scenarios, we isolate the accuracy impact of pruning from inherent circuit limitations. The goal is to determine whether the KAN can correct for pruning-induced errors and under what conditions this compensation remains effective.}

\blue{Algorithm~\ref{alg:akan_pruning} summarizes the complete 
hardware-aware pruning and training procedure. The four phases---SPICE characterization, coefficient-level pruning enumeration, per-configuration error modeling, and KAN training with error injection---are described in detail in the following subsections.}

\blue{
\begin{algorithm}[t]
\caption{Hardware-Aware Pruning and Training of AKAN}
\label{alg:akan_pruning}
\begin{algorithmic}[1]
\Require Target function $f_{\text{ideal}}(x)$, parameter grid $\{k_0,k_1,k_2\}\in[-0.5,0.5]^3$ (10 levels each), pruning configurations $\mathcal{P}$, threshold $R^2_{\min}=0.95$
\Ensure Trained KAN weights $\Theta^*$, best pruning config $p^*$
\State \textbf{// Phase 1: SPICE Characterization (run once)}
\For{each $(k_0,k_1,k_2)$ combination (1,000 total)}
    \State Run Cadence Spectre transient simulation
    \State Record $V_{\text{out}}(x_i)$ for $x_i \in [-0.5,0.5]$\,V
    \State Compute NMPE, area, power
\EndFor
\State Select top-30 simulations (NMPE closest to zero)
\State \textbf{// Phase 2: Coefficient-Level Pruning Enumeration}
\State $\mathcal{P} \leftarrow \{\text{baseline, prune }k_0,k_1,k_2, k_0k_1, k_0k_2, k_1k_2, k_0k_1k_2\} \times 3\text{ splines} \rightarrow 20\text{ unique configs}$
\For{each pruning config $p \in \mathcal{P}$}
    \State \textbf{// Phase 3: Error Modeling per Configuration}
    \State Compute $\epsilon_{\text{raw}}(x_i)$ from stored SPICE data with pruned coefficients set to zero
    \State Fit $\hat{\epsilon}_p(x)$ using polynomial/exp/log/rational candidates
    \State Select model with $R^2 \geq R^2_{\min}$ and fewest parameters
    \State Record hardware cost: $\text{area}(p)$, $\text{power}(p)$
    \State \textbf{// Phase 4: KAN Training with Error Injection}
    \State Set noisy target: $f_{\text{target}}(x) = f_{\text{ideal}}(x) + \hat{\epsilon}_p(x)$
    \State Train KAN via backpropagation on $\{(x_i, f_{\text{target}}(x_i))\}$
    \State Evaluate $\text{NMPE}(p)$ on validation set
\EndFor
\State $p^* \leftarrow \arg\min_{p} \text{NMPE}(p)$ subject to hardware budget
\end{algorithmic}
\end{algorithm}
}

\begin{figure}
\centering
\includegraphics[width=0.75\linewidth]{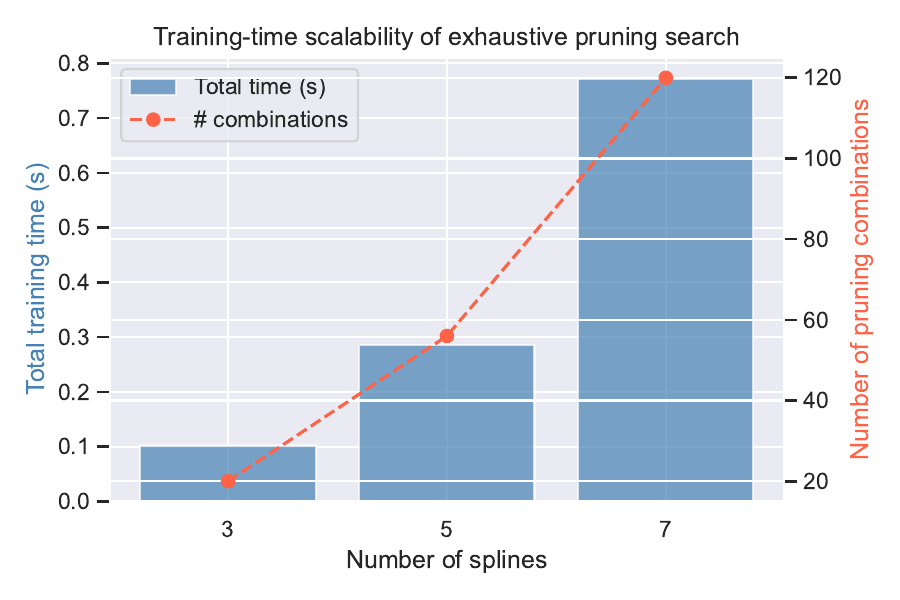}
\caption{Computational cost of the exhaustive pruning search for different AKAN configurations. The execution time remains under 1s for networks with up to 7 splines, confirming the efficiency of the analytical error model evaluation.}
\label{fig:scalability}\vspace{-2ex}
\end{figure}

The exhaustive search is computationally feasible because: (1) error model evaluation in Python is fast (milliseconds per configuration) compared to the initial SPICE characterization; (2) the 1,000 circuit simulations are performed once and reused across all KAN training experiments, amortizing the computational cost. \blue{Fig.~\ref{fig:scalability} shows the training-time scalability of the exhaustive pruning search for AKAN configurations with 3, 5, and 7 splines. Total wall-clock time remains under one second even for the 7-spline case, confirming the practical feasibility of the exhaustive search. For significantly larger networks, the framework can transition to a greedy or sensitivity-guided pruning strategy.}

\subsection{Validation and Performance Evaluation}

Validation is carried out by training AKAN models with both noiseless (ideal) and noisy (applying the error modeling) splines across all possible pruning configurations. \blue{In all experiments, the AKAN consists of a single hidden layer with 3 splines, giving a $[1 \to 3 \to 1]$ topology for single-input single-output approximation. This choice is motivated by the hardware constraints of FE: each additional spline incurs a significant area and power penalty in IGZO technology, making deeper or wider configurations infeasible within the target power budget.} Performance is quantified using the Normalized Mean Percentage Error (NMPE), enabling direct comparison of pruning strategies.

\blue{The six mathematical functions used for validation---$x^2$, $e^x$, $\sin(x)$, $\ln(x)$, $\tanh(x)$, and $\sqrt{x}$---are the \emph{target approximation functions} for each dataset, not an architectural constraint of the AKAN. The hardware architecture is a general-purpose analog function approximator capable of representing any sufficiently smooth function over $[-0.5\,\text{V}, +0.5\,\text{V}]$; the KAN is trained separately for each target function.}

\green{Beyond synthetic benchmarks, we validate the framework on four sensor datasets representative of FE deployment scenarios. This evaluation demonstrates not only the feasibility of hardware implementation but also the practical viability of pruned AKANs as low-power, compact accelerators for analog preprocessing in wearable and IoT applications.}

%% file: FigTab/flowchart.tex
\begin{figure}
\centering
\includegraphics[width=1\linewidth]{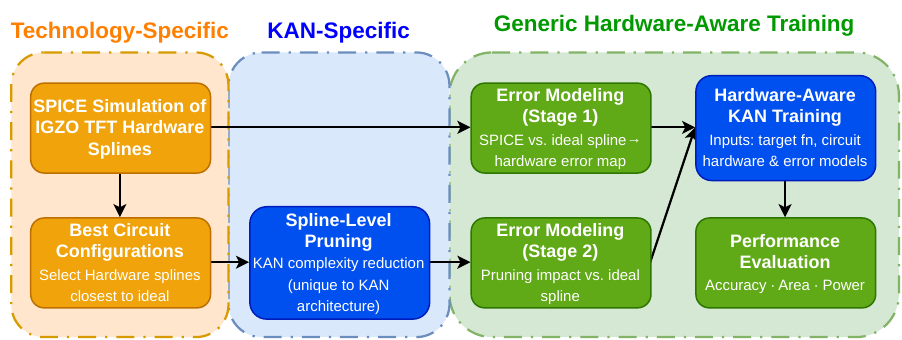}\vspace{-2ex}
\caption{Layered workflow of the proposed hardware-aware function approximation framework using Kolmogorov–Arnold Networks (KANs). 
The methodology separates technology-specific hardware characterization, KAN-specific pruning, and a generic hardware-aware training loop. 
SPICE simulations generate hardware spline models, error modeling captures deviations from ideal behavior, and training evaluates performance in terms of accuracy, area, and power.}
\label{fig:flowchart}\vspace{-2ex}
\end{figure}

%% file: FigTab/pruning.tex
\begin{figure}[t]
\centering
\includegraphics[width=0.9\linewidth]{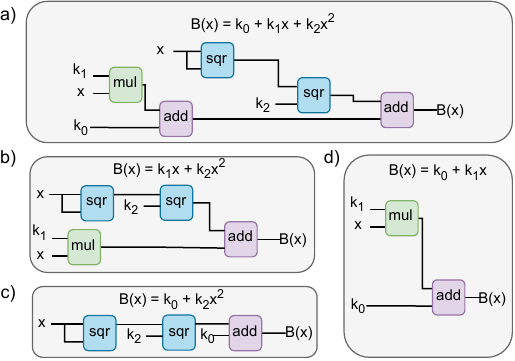}
\caption{Architecture of the spline: a) Original, and after pruning: b) \( k_0 \), c) \( k_1 \), and d) \( k_2 \).}
\label{fig:pruning}\vspace{-2ex}
\end{figure}

%% file: Section/4_Evaluation.tex
\section{Results and Analysis}\label{sec:results}

\begin{table*}
\centering
\caption{Detailed Simulation Setup\blue{. The AKAN topology used in all experiments is $[1\to3\to1]$ (one hidden layer with 3 splines). The six mathematical functions listed are the validation targets.}}
\scalebox{1}{\input{FigTab/simulationsetup}}\label{tab:simuationsetup}\vspace{-2ex}
\end{table*}

\subsection{Simulation Setup}

\autoref{tab:simuationsetup} summarizes the key parameters and tools used in the simulation and pruning process for the AKAN hardware architecture. It presents the simulation environment, parameter ranges, and evaluation conditions used to ensure consistent and reproducible hardware characterization across all pruning experiments.
\textcolor{black}{
All simulations reported in this work are performed at SPICE level using the Cadence Spectre simulator, as detailed above. Importantly, after training, the learned spline coefficients are re-inserted into the circuit netlist, and the final NMPE values reported in this section are computed from independent post-training SPICE simulations.}
\textcolor{black}{All area and power figures reported in this section are scoped to the AKAN analog preprocessing block. The system-level significance of these improvements depends on the fraction of the complete sensing pipeline occupied by this block.}

\subsection{\blue{PVT Robustness Analysis}}

\blue{To assess the practical reliability of the AKAN under IGZO-relevant non-idealities, we conducted a comprehensive PVT characterization across 27 operating conditions: 3 process corners (SS, TT, FF) $\times$ 3 supply voltages (0.9\,V, 1.0\,V, 1.1\,V) $\times$ 3 temperatures (0°C, 27°C, 40°C), covering the full wearable healthcare operating range~\cite{lee:flexiblepatch, wearableskinpatch, Haghi:wearableinhealthmonitoring2021}.

\begin{figure}
\centering
\includegraphics[width=1\linewidth]{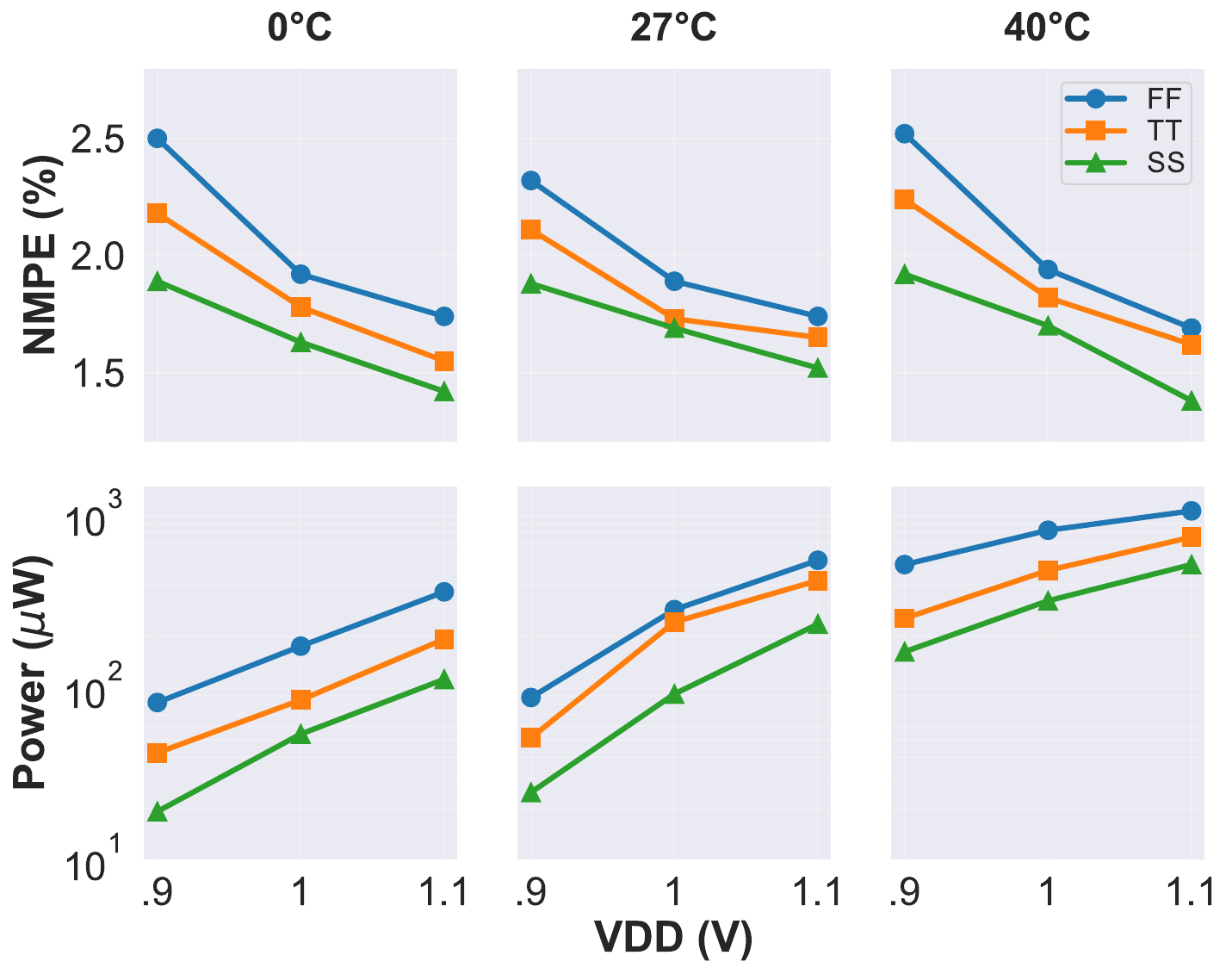}\vspace{-2ex}
\caption{System-level robustness analysis across 27 PVT corners ($V_{DD} \in \{0.9, 1.0, 1.1\}$~V, Temp. $\in \{0, 27, 40\}^{\circ}$C, and FF/TT/SS process corners). The top row illustrates NMPE stability, demonstrating hardware-aware compensation, while the bottom row shows the total power consumption on a log scale.}\vspace{-2ex}
\label{fig:pvt_analysis}
\end{figure}

\textcolor{black}{The PVT characterization presented here covers systematic process, voltage, and temperature variations through 27 corner simulations. 
However, it does not capture the effect of local device mismatch, which arises from random spatial variation in $V_{\text{TH}}$ and carrier mobility and can independently perturb the input--output behavior of individual spline circuits. 
A full Monte Carlo mismatch and yield analysis is an important next step for validating the AKAN in a production-intent FE context and is a direction for future work.}
The system-level robustness of the AKAN spline across 27 distinct PVT operating conditions is summarized in Fig.~\ref{fig:pvt_analysis}. As observed in the top panels, the NMPE remains remarkably stable within the wearable temperature range (27--40$^{\circ}$C), with a maximum deviation ($\Delta$NMPE) of only 0.74\,pp relative to nominal conditions. Even at the 0$^{\circ}$C extreme, the worst-case NMPE is contained at 2.50\%, confirming that the KAN training framework effectively compensates for corner-induced shifts by adapting the spline weights. \blue{These results indicate that the derived error model captures the dominant systematic hardware behavior across the evaluated environmental conditions.}

Regarding power stability (Fig.~\ref{fig:pvt_analysis}, bottom), the AKAN exhibits the expected leakage increase at elevated temperatures---a characteristic of unipolar IGZO TFTs. However, the nominal power consumption (TT, 1.0\,V, 27$^{\circ}$C) remains highly efficient at 241.8\,$\mu$W. By leveraging the proposed coefficient-level pruning, this power footprint can be further reduced by up to 50\% depending on the target function, maintaining a competitive power-to-performance ratio suitable for long-term wearable monitoring.

Regarding mechanical bending variability: the current PDK does not include explicit compact models for bending stress. Based on published characterization of this specific process~\cite{EuropracticeFlexibleElectronics}, bending typically induces shifts in threshold voltage ($V_{\text{TH}}$) and carrier mobility. Our framework is inherently well-suited to compensate for such parameter shifts, since the hardware error model $\hat{\epsilon}_{\text{hardware}}(x)$ treats them as systematic additive errors identical in form to those captured in the PVT sweep. 
Embedding a bending-corner error model into the training loop using the same pipeline described in Section~\ref{sec:methodology} is therefore a straightforward extension of the present framework.}

\subsection{Analysis of NMPE and Parameter Distributions for Top Simulations}

\input{FigTab/NMPEdistribution}

\textcolor{black}{To assess the sensitivity of the representative error model to the selection threshold, representative curves were derived using Top-10, Top-20, Top-30, Top-50, and Top-100 configurations ranked by NMPE. The resulting curves showed maximum deviations below 27mV relative to the Top-30 model for all subsets up to Top-100, confirming that the proposed error model is robust to the exact subset size.}
\textcolor{black}{
To further validate the representativeness of the proposed error model, we analyze both the full design-space distribution (1,000 SPICE simulations) and the calibrated operating regime defined by the top-30 configurations selected according to NMPE. 
While the full design space exhibits large structural variability due to differences in spline coefficient parameterizations, the calibrated subset shows significantly reduced variability and concentrates around the proposed representative model. 
In particular, the 5th–95th percentile envelope of the calibrated subset yields a maximum spread of 189.82 mV across the input range, confirming that the proposed model accurately captures the dominant behavior of hardware instances within the operational regime.}
In Fig.~\ref{fig:NMPEdistribution}, the first graph illustrates the NMPE distribution for the 30 best simulations, where the most accurate ones are those with an NMPE closest to zero, either positively or negatively\green{. It is important to note that} a negative NMPE indicates that the approximation tends to underestimate the function, while a positive value suggests an overestimation. 
The results show that all 30 simulations fall within a range of -13\% to 13\%, with more than 50\% of them concentrated between -5\% and 5\%. Previous studies have shown that an error of -7\% has minimal impact when integrated into the KAN. \green{For the applications considered in this work---biosignal preprocessing and classification---both positive and negative errors within $\pm5$\% produce negligible degradation in downstream task performance, as validated in~\autoref{sec:results}.} However, the effect of values approaching $\pm13$\% remains to be examined \green{and represents the operational boundary of acceptable accuracy for the proposed architecture}.

The remaining graphs present the distribution of \( k_0 \), \( k_1 \), and \( k_2 \) for these top-performing simulations. Several observations can be drawn: 
First, it is reasonable that \( k_0 \) tends to be positive and relatively large, as it is the only term independent of the input \( V_{in} \) and acts solely as an additive factor. The design appears more robust at higher positive voltages, an intuitive result confirmed by the simulations. 
Regarding \( k_1 \) and \( k_2 \), values close to zero do not fall among the most optimal values for getting a good NMPE, as both terms are multiplied by the input signal.
\green{From a hardware perspective, near-zero coefficients represent circuit configurations where transistors operate near threshold or in weak inversion regions characterized by high sensitivity to process variations and temperature drift. 
\blue{Pruning these parameters reduces circuit complexity while removing operating points that are more sensitive to mismatch and noise, explaining the simultaneous improvement in accuracy and power efficiency.}}
This distribution analysis directly informs the pruning thresholds applied in the following section, establishing bounds based on empirical parameter importance rather than arbitrary cutoffs.

\orange{
\subsection{Evaluation Scenarios}

We define three evaluation scenarios to isolate different sources of error:
\begin{itemize}
    \item \textbf{Ideal (noiseless):} Training assumes perfect mathematical spline behavior with no hardware imperfections. This represents an upper bound on achievable accuracy and serves as a reference for ideal performance.
    \item \textbf{Baseline (noisy, unpruned):} Training incorporates the measured hardware error from the unpruned circuit. This isolates the inherent circuit limitations (mismatch, nonlinearity, noise) from pruning effects, establishing a realistic baseline for hardware-implemented AKANs.
    \item \textbf{Pruned (noisy):} Training incorporates hardware error from pruned configurations. Performance relative to the baseline scenario reveals the isolated impact of parameter removal, allowing us to quantify the accuracy-cost trade-off of pruning.
\end{itemize}

This decomposition allows us to separately quantify: (1) how much error is due to analog circuit imperfections in the best-case (unpruned) scenario, and (2) how much additional error is introduced specifically by removing polynomial coefficients through pruning.
}

\subsection{Impact of Pruning on KAN Model Accuracy and Efficiency}

\begin{table}
\centering
\footnotesize
\caption{Hardware cost of an analog pruned spline across all the architectures presented in Fig.~\ref{fig:implementation}}
\scalebox{1}{\input{FigTab/hardwarecost}}
\begin{tablenotes}
\footnotesize
\item $^1$ Area Reduction. $^2$ Power Reduction
\end{tablenotes}
\vspace{-3ex}\label{tab:prunedcost}
\end{table}

By analyzing the distribution of parameters \( k_0 \), \( k_1 \), and \( k_2 \), we can determine the optimal values for pruning. In this work, we selected a pruning range of \([-0.06, 0.06]\) for \( k_0 \) and \([-0.17, 0.17]\) for \( k_1 \) and \( k_2 \). \green{These thresholds were selected based on the parameter distribution analysis shown in Fig.~\ref{fig:NMPEdistribution}: coefficients within these ranges correspond to the tails of the distribution where individual parameters contribute minimally to overall approximation accuracy.} \blue{These thresholds prioritize the removal of parameters with minimal contribution to the function approximation while preserving dominant functional components.}
\purple{In our SPICE simulations, when a parameter falls within these thresholds, we manually set its contribution to zero and remeasure the error, modeling its effect accordingly.}
The hardware cost associated with each pruning strategy is presented in~\autoref{tab:prunedcost}, where it can be observed that an optimal pruning of \( k_1 \) or \( k_2 \) can lead to area reductions between 45\% and 55\%, along with a 50\% decrease in power consumption. 
\blue{These reductions correspond to smaller circuit footprint and lower energy budgets, which are critical design constraints for FE implementations.}

\subsection{Validation on Representative Datasets}
\input{FigTab/resultsECG}
\begin{figure}
\centering
\includegraphics[width=1\linewidth]{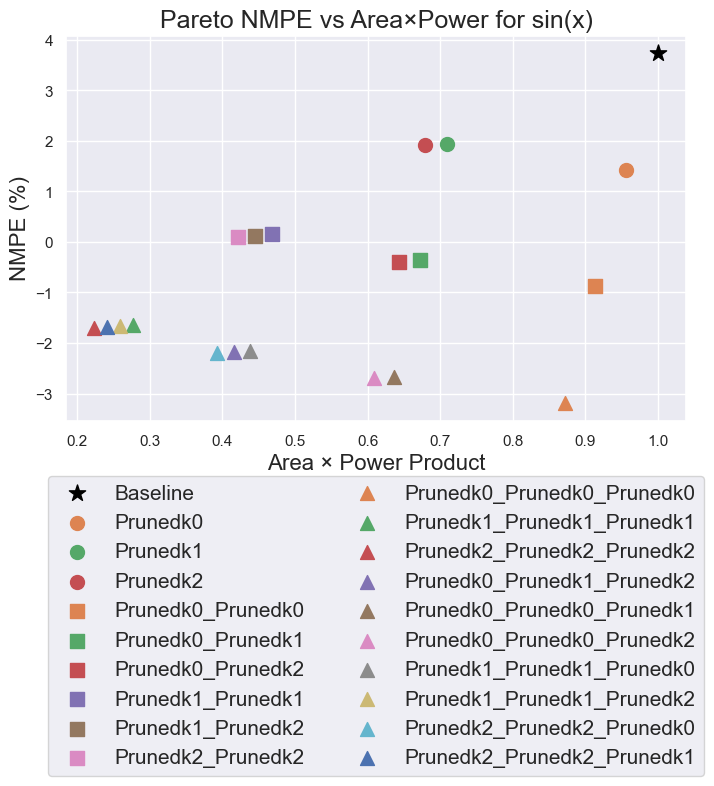}\vspace{-2ex}
\caption{Pareto distribution of NMPE versus $Area \times Power $ for the Iris Petal Length dataset approximating $\sin(x)$. The plot shows the baseline (3 non-pruned splines, star marker) and all possible pruning scenarios across the three splines in the network. Pruning configurations are labeled according to which coefficient ($k_0$, $k_1$, or $k_2$) is removed from each spline. Multiple configurations achieve lower NMPE and reduced $Area \times Power$ compared to the baseline, demonstrating favorable Pareto improvement through strategic parameter removal.}
\label{fig:pareto}
\vspace{-2ex}
\end{figure}

Our proposed AKAN framework is validated with four representative datasets\blue{, each evaluated using an AKAN of topology $[1\to3\to1]$}.
\blue{These datasets were selected to represent diverse sensing modalities commonly encountered in flexible and wearable electronic systems.}
\begin{itemize}
    \item \textbf{PPG-DaLiA \cite{ppg-dalia}:} Photoplethysmogram (PPG) signals, suitable for wearable health monitoring and stress detection applications.  
    \item \textbf{ECG5000 \cite{ecg5000}:} Electrocardiogram (ECG) time series, relevant for cardiac anomaly detection and biomedical signal processing.  
    \item \textbf{Household Electric Power Consumption \cite{individual_household_electric_power_consumption_235}:} Domestic energy usage data, useful for temporal function approximation in energy analytics.  
    \item \textbf{Iris dataset \cite{iris_53}:} A classical benchmark for multi-class classification, demonstrating the ability of AKAN to approximate decision functions.  
\end{itemize}
Preprocessing ensures compatibility with hardware constraints: inputs are normalized and mapped to the voltage range $[-0.5, 0.5]$ V, while target values are paired with corresponding mathematical functions or decision boundaries. 
\green{The choice of mathematical functions---$\tanh(x)$, $\exp(x)$, $\ln(x)$, $\sin(x)$, $\sqrt{x}$, and $x^2$---is motivated by their prevalence in analog signal conditioning pipelines~\cite{Zikang:analogfrontendsflexible} and their representation of fundamental nonlinear operations in sensor interfaces~\cite{yang2025efficient}.}
Specifically: (1) Hyperbolic tangent $\tanh(x)$ models saturation characteristics in bioamplifiers~\cite{HarrisonAfeBioamp2003}; (2) Exponential $\exp(x)$ appears in temperature compensation circuits and logarithmic ADCs~\cite{lee:flexiblepatch}; (3) Logarithmic $\ln(x)$ enables dynamic range compression for photodetectors and acoustic sensors~\cite{SedraSmithMicroelectronics}; (4) Trigonometric $\sin(x)$ captures periodic physiological rhythms~\cite{yang2025efficient}; (5) Square root $\sqrt{x}$ implements RMS computation for power monitoring~\cite{AnalogDevicesRMS}; (6) Quadratic $x^2$ enables polynomial calibration in resistive sensors~\cite{Zikang:analogfrontendsflexible}. 
\green{For each dataset, the pruning strategy was applied across three spline types and 20 unique pruning combinations, with averages reported in~\autoref{tab:ECG_results}.}

\begin{table}
\centering
\caption{Downstream application metrics: classification accuracy and F1-score.  
``Ideal AKAN'' includes the analog hardware but assumes perfect (noiseless) circuit behavior. 
``HW Baseline'' is the unpruned analog AKAN with measured hardware errors.}
\label{tab:downstream_acc}
\begin{tabular}{lcccc}
\toprule
\textbf{Config} & \multicolumn{2}{c}{\textbf{ECG5000}} & \multicolumn{2}{c}{\textbf{Iris}} \\
& Acc (\%) & F1 (\%) & Acc (\%) & F1 (\%) \\
\midrule
Ideal AKAN (noiseless) & 58.5 & 73.8 & 95.3 & 95.3 \\
HW Baseline (unpruned)     & 58.4 & 73.7 & 95.3 & 95.4 \\
\textbf{Best Pruned AKAN}       & \textbf{58.4} & \textbf{73.7} & \textbf{95.3} & \textbf{95.3} \\
\bottomrule
\end{tabular}
\end{table}

\begin{table}
\centering
\caption{Downstream signal integrity: NRMSE for regression tasks. 
``HW Baseline'' is the unpruned analog AKAN with measured hardware errors.}
\label{tab:downstream_nrmse}
\begin{tabular}{lcc}
\toprule
\textbf{Dataset} & \textbf{Baseline NRMSE (\%)} & \textbf{Best Pruned} \\
&  & \textbf{NRMSE (\%)}  \\
\midrule
IHEPC (IoT Energy)   & 7.77  & 5.23 \\
PPG-DaLiA (Acc-X)    & 31.59 & 24.03 \\
PPG-DaLiA (Acc-Y)    & 35.65 & 26.17 \\
\bottomrule
\end{tabular}
\end{table}
\blue{Downstream application-level evaluation: To move beyond NMPE as a proxy metric, we evaluated the AKAN across three downstream scenarios. For classification (ECG5000, Iris), we report accuracy and F1-score; for signal reconstruction (IHEPC, PPG-DaLiA), we report NRMSE. Results are summarized in Tables~\ref{tab:downstream_acc} and~\ref{tab:downstream_nrmse}.}

\blue{The pruned AKAN matches the software-only KAN on both classification tasks (58.4\% accuracy / 73.7\% F1 on ECG5000; 95.3\% on Iris), confirming that analog hardware non-idealities and pruning do not degrade downstream task performance. For ECG5000, the F1-score is the more informative metric given the inherent 5-class imbalance of this dataset. For signal reconstruction, the Best Pruned AKAN outperforms the unpruned hardware baseline on all datasets, with NRMSE reductions of up to 33\%. Near-zero coefficients map to transistors biased near threshold, where sensitivity to process variation is highest. Removing these terms reduces the dominant source of systematic output error, improving reconstruction accuracy while simultaneously reducing area and power.}

\blue{For hardware context, the closest complete FE inference system~\cite{FlexCoProc:DATE26}---a flexible RISC-V with bespoke MAC co-processor for MLP inference---achieves 92.84\% ECG5000 accuracy at 2.42\,mm$^2$ and 1.519\,mW. The pruned AKAN, as a programmable analog preprocessing block, operates at 0.096\,mm$^2$ and 0.504\,mW on average, offering a $25\times$ area and $3\times$ power reduction relative to that complete system. These two works target different stages of the sensing pipeline and are complementary.}
\green{This regularization effect can be understood through the lens of model complexity: unpruned splines with three independent coefficients possess sufficient degrees of freedom to overfit the training voltage range. By constraining the parameter space through systematic removal of near-zero coefficients, the pruned model is forced to represent only the dominant functional characteristics, inherently improving generalization.}

\green{The effectiveness of pruning is dataset-dependent. While most cases show clear gains (e.g., ECG5000 and Iris datasets), some configurations degrade slightly, as seen in certain Household Power scenarios. }
\green{Analysis reveals that degradation occurs primarily when: (1) the target function exhibits high curvature requiring all three polynomial terms for accurate representation (e.g., $\exp(x)$ near saturation), or (2) the input data distribution is highly non-uniform, concentrating samples in regions where pruned terms would otherwise contribute significantly. 
For the Household Power dataset approximating $\ln(x)$, 73\% of input samples fall below -0.3\,V where the $k_2$ term dominates; pruning this coefficient therefore has a disproportionate impact on approximation accuracy.
Conversely, ECG5000 signals are more uniformly distributed across $[-0.5, 0.5]$\,V, allowing aggressive pruning without performance loss.}

\green{From a hardware perspective, the gains are substantial: pruning reduces area by 29.8\% on average (up to 55\%) and power by 30.5\% (up to 50.3\%). 
Fig.~\ref{fig:pareto} illustrates this trade-off for the Iris Petal Length dataset\orange{, where the target function is $\sin(x)$}.
Multiple pruning scenarios achieve lower NMPE than the baseline while simultaneously reducing $Area \times Power$, confirming that the Pareto frontier shifts favorably with pruning.}

The consistent improvements across physiological, energy, and classification tasks demonstrate that pruned AKANs remain robust under these conditions, supporting their viability for practical FE-based sensor preprocessing in wearable and IoT applications.

%% file: FigTab/simulationsetup.tex
\begin{tblr}{
  colspec = {Q[110] Q[150] Q[220]}, 
  vline{2,3} = {-}{0.4pt}, 
  hline{1,2,15,16,21} = {1.2pt}, 
  hline{3,4,5,6,7,8,9,10,11,12,13,14,17,18,19,20} = {2,3}{0.4pt}, 
}

\textbf{Simulation Parameter} & \textbf{Value} & \textbf{Description} \\

\textbf{Technology} & \blue{PragmatIC FlexICs PDK (Helvellyn v2.1.0)} & \blue{Subtractive flexible electronic technology used for circuit simulations.} \\
\textbf{Simulation Tool} & \blue{Cadence Spectre (SPICE simulator)} & \blue{Used for transistor-level circuit simulations.} \\
\textbf{Operating Voltage} & 1.0V (V$_{dd}$), -1.0V (V$_{ss}$) & Supply and ground voltages. \\
\textbf{Operating Temperature} & \blue{27$^\circ$C} & \blue{Nominal operating temperature used for baseline simulations.} \\
\textbf{Input Signal} & Ramp from -0.5V to 0.5V & The ramp signal used for input (V$_{in}$). \\

\textbf{Parameter Sweep} & 
\blue{$k_0, k_1, k_2$ each evaluated at 10 discrete values within [-0.5V, 0.5V]} & 
\blue{Systematic parameter exploration for robustness and error characterization.} \\

\textbf{Number of Simulations} & 1,000 & 
\blue{Total number of parameter combinations evaluated from the coefficient sweep.} \\

\textbf{Sampling Interval} & 0.02V & 
\blue{Input voltage step size used to sample the transfer characteristic.} \\

\textbf{Software Environment} & 
\blue{Python 3.9, Jupyter Notebook, Anaconda} & 
\blue{Environment used for data processing, model training, and visualization.} \\

\textbf{Libraries Used} & 
\blue{NumPy, Pandas, SciPy, Matplotlib, Seaborn} & 
\blue{Libraries used for numerical analysis, visualization, and curve fitting.} \\

\textbf{Mathematical Functions} & 
\(x^2\), \(e^x\), \(\sin(x)\), \(\ln(x)\), \(\tanh(x)\), \(\sqrt{x}\) & 
\blue{For \(\ln(x)\) and \(\sqrt{x}\), the input domain was shifted by +0.5 to ensure strictly positive arguments.} \\

\textbf{Function Fitting} & 
SciPy curve fitting & 
\blue{Used to model hardware-induced error functions from simulation data.} \\

\textbf{KAN Architecture} & 
\blue{3 splines} & 
\blue{Number of spline units composing the Kolmogorov-Arnold Network.} \\

\end{tblr}

%% file: FigTab/NMPEdistribution.tex
\begin{figure}
\centering
\includegraphics[width=1\linewidth]{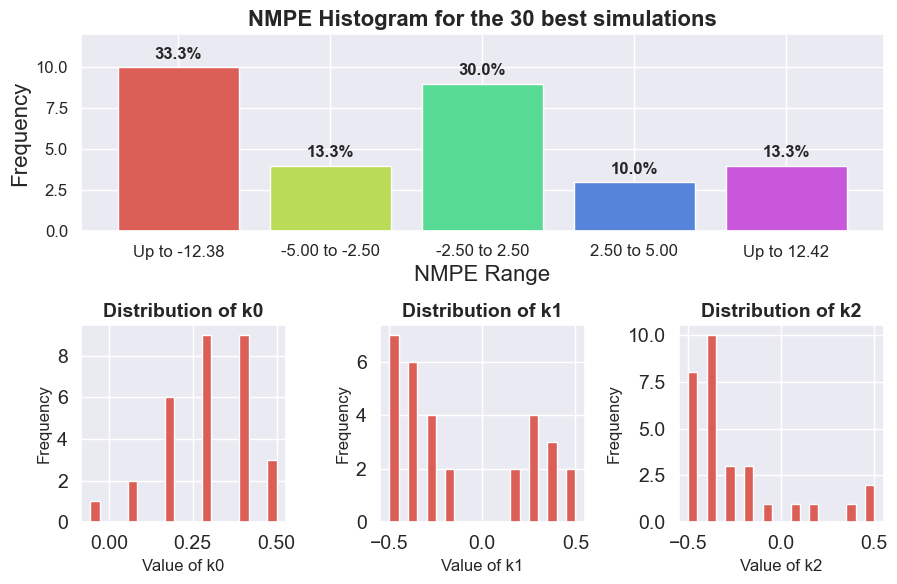}
\caption{Distributions of NMPE and Parameters \( k_0 \), \( k_1 \), and \( k_2 \) for the Top 30 Simulations}
\label{fig:NMPEdistribution}
\vspace{-2ex}
\end{figure}

%% file: FigTab/hardwarecost.tex
\setlength{\arrayrulewidth}{0.4pt} 

\begin{tblr}{
  colspec = {Q[100] Q[80] Q[80] Q[50] Q[50]},  
  vline{2,3,4,5} = {-}{0.4pt}, 
  hline{2,6} = {1.2pt}, 
  hline{1} = {2,3,4,5}{1.2pt}, 
}
    & \textbf{Area($\mu m^2$)} & \textbf{Power($\mu W$)} & \textbf{AR$^1$} & \textbf{PR$^2$} \\ 
    \textbf{Original} & 45570 & 241.8 & -- & -- \\
    \textbf{Pruning \( k_0 \)} & 43613 & 220.2 & 4.3\% & 8.9\% \\
    \textbf{Pruning \( k_1 \)} & 25035 & 121.9 & 45.1\% & 49.6\% \\
    \textbf{Pruning \( k_2 \)} & 20536 & 120   & 55\%   & 50.4\% \\
\end{tblr}

%% file: FigTab/resultsECG.tex
\begin{table*}[t]
\centering
\footnotesize
\caption{Summary of pruning impact on average and best NMPE across datasets, together with average area and power reduction.}
\label{tab:ECG_results}

\begin{threeparttable}
\begin{tblr}{
  colspec = {Q[35] Q[70] Q[70] Q[70] Q[40] Q[40] Q[40]},
  hline{1,2,15} = {1.2pt},
  rowsep = 1.6pt, 
  hline{3,7,8,12,13,14} = {0.8pt},
  column{2-7} = {c},
}
\textbf{Dataset} & 
\textbf{Avg. NMPE (Base)$^{1}$} & 
\textbf{Best NMPE (AP)$^{2}$} & 
\textbf{Avg. NMPE (AP)$^{3}$} & 
\textbf{$\Delta$NMPE$^{4}$} & 
\textbf{Avg. AR(\%)$^{5}$} & 
\textbf{Avg. PR(\%)$^{6}$} \\ 
\SetCell[c=7]{l} \textbf{PPG-DaLiA} \\
PPG Chest & 0.26 & 0.01 & -0.22 & 0.04 & \SetCell[r=4]{c} 29.8 & \SetCell[r=4]{c} 30.5 \\
PPG Wrist & 0.19 & -0.01 & -0.24 & -0.05 &  &  \\
Acc X     & -0.91 & -0.05 & -1.35 & -0.4 &  &  \\
Acc Y     & 1.64 & 0.04 & 1.20 & 0.44 &  &  \\
\SetCell[c=7]{l} \textbf{Iris} \\
Sepal L. & 3.08 & -0.08 & -2.46 & 0.62 & \SetCell[r=4]{c} 29.8 & \SetCell[r=4]{c} 30.5 \\
Sepal W. & 4.90 & -0.05 & -1.36 & 3.24 &  &  \\
Petal L. & 3.77 & -0.07 & -1.89 & 1.88 &  &  \\
Petal W. & 4.77 & -0.06 & -1.59 & 3.18 &  &  \\ 
\textbf{ECG5000}   & 1.73 & 0.04 & 1.24 & 0.49 & 29.8 & 30.5 \\
\textbf{IHEPC$^{7}$} & -2.37 & -0.16 & -4.70 & -1.97 & 29.8 & 30.5 \\
\textbf{Average} & 1.7 & -0.04 & -1.09 & 0.75 & 29.8 & 30.5 \\
\end{tblr}

\begin{tablenotes}
\footnotesize
\item $^{1}$ NMPE of the baseline configuration (non-pruned). $^{2}$ Best NMPE obtained among all pruning configurations (AP = After Pruning). $^{3}$ Average NMPE across all pruning configurations. $^{4}$ Signed change in NMPE, where a positive value indicates that pruning moves the NMPE closer to zero, and a negative value indicates degradation. $^{5}$ Average Area Reduction (\%). $^{6}$ Average Power Reduction (\%). $^{7}$ IHEPC = Individual Household Electric Power Consumption dataset.
\end{tablenotes}
\end{threeparttable}\vspace{-2ex}
\end{table*}

%% file: Section/5_Conclusion.tex
\section{Conclusion}\label{sec:conclusion}
This work explores the impact of pruning strategies on the efficiency of Analog Kolmogorov-Arnold Networks (AKANs) for function approximation in Flexible Electronics (FE). By systematically pruning the circuit design, we demonstrate that substantial reductions in hardware cost can be achieved while simultaneously improving approximation accuracy.
The results show that pruning reduces average hardware cost by approximately 30\% in both area and power, with best-case reductions reaching 55\% in area and 50\% in power. At the same time, NMPE consistently improves compared to the unpruned baseline, indicating that sparsification of spline coefficients acts as an implicit regularizer for the analog function representation.
\textcolor{black}{The reported area and power gains are evaluated at the AKAN preprocessing block level for a $[1\!\to\!3\!\to\!1]$ network, with full PPA characterization across all pruning configurations.}
\blue{At the system level, pruned AKANs maintain classification performance (accuracy and F1-score) comparable to the unpruned baseline, while improving signal reconstruction quality. Robustness analysis across 27 PVT operating corners further shows stable NMPE behavior within a 0.74\,pp variation in the wearable temperature range, confirming reliable operation under realistic IGZO process and environmental conditions.}
These findings underscore the potential of pruned AKANs for function approximation in FE, demonstrating that strategic parameter reduction can simultaneously enhance both power and area efficiency while maintaining or improving accuracy. This strategy is particularly advantageous for resource-constrained applications, where minimizing hardware overhead is paramount while preserving computational performance.